\documentclass[twocolumn,showpacs,amsmath,amssymb,pre,aps]{revtex4}   
\usepackage{graphicx}
\usepackage{dcolumn}
\usepackage{bm}

\begin{document}
\title{
Unconventional Vortex Dynamics in Mesoscopic Superconducting Corbino Disks
}
\draft

\author{
N.S.~Lin, V.R.~Misko, and F.M.~Peeters
}
\affiliation{
Department of Physics, University of Antwerpen, Groenenborgerlaan 171, B-2020 Antwerpen, 
Belgium
}

\date{\today}

\begin{abstract}
The discrete shell structure of vortex matter strongly influences the flux dynamics 
in mesoscopic superconducting Corbino disks. While the dynamical behavior is 
well understood in large and in very small disks, in the intermediate-size regime it 
occurs to be much more complex and unusual, due to (in)commensurability 
between the vortex shells. We demonstrate unconventional vortex dynamics 
(inversion of shell velocities with respect to the gradient driving force) and angular 
melting (propagating from the boundary where the shear stress is minimum, 
towards the center) in mesoscopic Corbino disks.
\end{abstract}
\maketitle

Investigation of mesoscopic and nano-structured superconductors has become one of the major 
topics of experimental and theoretical studies on vortex matter in superconductors over the 
last two decades. 
The unique possibility to investigate the interplay between the vortex-vortex and vortex-boundary 
interactions, the size-, shape-, and symmetry-induced effects makes mesoscopic superconductors 
very attractive from both the fundamental research and possible fluxonics  applications, i.e., 
manipulating single flux quanta similar to electrons in nano- and microelectronics. 
Thus it was demonstrated theoretically 
\cite{DSPGL97,SPB98,SPL99,BPS01}
and experimentally 
\cite{GPN97,kanda,grigorieva,irina07} 
that in mesoscopic superconducting disks shape effect leads to the formation of circular-symmetric 
giant vortex states 
\cite{GPN97,kanda}
or concentric shells of vortices 
\cite{grigorieva,slava_disk}. 
Furthermore, symmetry-induced vortex-antivortex ``molecules'' were theoretically predicted \cite{vav} 
in mesoscopic squares and triangles. 
On the other hand, the dynamics of vortex matter in disks can be studied using a Corbino setup. 
In the Corbino geometry an applied current produces a gradient in 
the Lorentz force, i.e., introduces a shear driving force between the rings (in large disks \cite{crabtree}) 
or shells (in mesoscopic regime \cite{wecorbino}) of vortices. 
This provides the unique opportunity to study various dynamical effects related to vortex motion, e.g., 
the onset of plasticity, 
channeling \cite{chan}, vortex friction \cite{fric}, etc. 
Moreover, the dynamics of self-organized vortex matter in mesoscopic disks has many common features 
to, e.g., atomic matter, charged particles in Coulomb crystals, vortices in rotating Bose-Einstein 
condensates, magnetic colloids, synthetic nanocrystals, etc. 
\cite{mitchell,pertsinidis,murray}, 
or charged balls diffusing in macroscopic Wigner rings 
\cite{coupier}. 

Several early studies on vortex matter in Corbino disks focused on the transition from elastic 
to plastic vortex motion 
\cite{crabtree,paltiel,marchetti}. 
Indeed, the radial current density $j$ in a Corbino disk decays as 
$I_{0}/r$ along the radius resulting in a stronger Lorentz force 
$F_{L}$ driving vortices near the disk center than near the edge. 
For small driving currents, the local shear stress is small and the whole vortex pattern, although 
elastically deformed, moves as a rigid body. 
For applied currents larger than some critical value $I_{c}$ 
a strong spatially inhomogeneous stress breaks up the vortex solid, and concentric 
annular regions (in large disks) or shells (in mesoscopic disks) move with different angular velocities. 
The different dynamical phases (e.g., elastic motion, shear-induced plastic slip) of vortex motion 
can be detected by measuring voltage profiles \cite{crabtree}. 
In large Corbino disks, the onset of plasticity was theoretically analyzed within a continuous 
model \cite{marchetti} and using molecular dynamics (MD) simulations of interacting vortices at $T=0$ 
\cite{miguel}. 
The magnetization and different scenarios of magnetic-flux penetration in a 
thin Corbino disk were studied \cite{clem} using the critical-state model. 
The dynamics of vortex shells in {\it small } mesoscopic Corbino disks containing up to 40 vortices 
(cf. Ref.~\cite{grigorieva}) arranged in two- and three-shell configurations, was studied using 
MD simulations for $T=0$ and $T>0$ \cite{wecorbino}. 
It was shown that the critical current $I_{c}$ had a non-monotonic dependence on the magnetic field $h$. 
The dynamical instabilities associated with a jump in $I_{c}(h)$ were revealed. 
The found unusual behavior was shown to be related to a ``structural transition'', i.e., an inter-shell 
vortex transition with increasing $h$. 
In the presence of pinning, thermally-activated externally-driven flux motion was analyzed for $T>0$. 
The investigation of the angular thermal melting explained the 
experimentally observed two-step melting transition in Corbino disks \cite{crabtree}. 

Here, we study the dynamics of vortex shells in larger mesoscopic  Corbino disks containing several shells. 
The essential {\it qualitative } difference to the earlier studied case \cite{wecorbino} is the following: 
While in small disks vortex shells are nearly perfectly circular, due to the strong confinement, in larger 
disks the vortex-vortex interaction becomes more important. 
The dynamics of vortex shells is influenced, to a great extent, by the interplay between the gradient 
Lorentz force and the (in)commensurability effects between the numbers of vortices in adjacent 
(and even more remote!) vortex shells. 
Note that all these factors are sensitive to the radius of the disk. 
As a result of the complex interplay of the above five factors, 
i.e., a) confinement, b) the inter-vortex interaction, c) the gradient Lorentz force, 
d) (in)commensurability, and e) the size of the disk, 
the system of vortex shells in large Corbino 
disks demonstrates very rich and {\it unusual} dynamical behavior. 
Thus we found: 
(i) unconventional angular melting, i.e., when the melting first occurs in regions 
where the shear stress is {\it minimum}, and 
(ii) unconventional shell dynamics when a shell that experiences a {\it weaker} Lorentz force moves 
{\it faster} than the adjacent shell driven by a stronger force. 
We expect that our results can be useful for understanding and predicting the dynamics 
of complex systems of interacting particles confined to an external potential such as colloids, 
vortices in Bose-Einstein condensates, charged particles, etc., or even for better understanding 
of mechanisms of motion of some biological objects.

{\it Theory and simulation.---}
We place a Corbino disk which has thickness $d$ and radius $R$ in 
a perpendicular external magnetic field $\bm{H}_0$. 
An external current flows radially from the center to the edge of the disk
and 
results in the inhomogeneous sheath current density 
$j(\rho) = I_{ext}/2\pi\rho d$. 
The Lorentz force (per unit length) acting on vortex $i$, $\Phi_0 \bm{j} \times \hat z$,
resulting from the external current is: 
\begin{equation}
\bm{f}^d_i = \frac{\Phi_0 I_{ext}}{2\pi d} \frac{\bm{\rho}_i \times
\hat z }{\rho_i^2} = f_{0} I_{0} \frac{\bm{r}_i \times \hat z}{r_i^2},
\label{eqfj}
\end{equation}
where $\bm{\rho}_i$ is the vortex position, 
$\bm{r}_i = \bm{\rho}_i/R$, 
and $\hat z$ is the unit vector along the magnetic field direction. 
Here the unit of force 
$f_0 = \Phi_0^2 / 2 \pi \mu_0 R \lambda^2 
= 4 \pi\mu_0\xi^2 H_c^2 / R$ and 
$I_{0} = \mu_0 \Lambda I_{ext} / \Phi_0$ 
are expressed in terms of 
the coherence length $\xi$, the magnetic field penetration depth $\lambda$ 
($\Lambda = \lambda^2 / d$ is the effective penetration depth in a superconductor 
of thickness $d$), 
and the thermodynamic $H_{c}$ and the upper $H_{c2}$ critical magnetic fields. 
In a thin superconducting disk such that $d < \xi \ll R \ll \Lambda$, 
the vortex-vortex interaction force 
$\bm{f}^{vv}_{i}$
and the force of the vortex interaction with the shielding currents and 
with the edge 
$\bm{f}^{s}_{i}$
can be modelled respectively by \cite{buzdin,CBPB04,BCPB04}
\begin{eqnarray}
\bm{f}^{vv}_{i} = f_{0} \sum \limits_{i,k}^{L} 
\left(\frac{\bm{r}_i-\bm{r}_k}{\left|\bm{r}_i
- \bm{r}_k\right|^2}- r_k^2\frac{r_k^2\bm{r}_i - \bm{r}_k}
{\left|r_k^2\bm{r}_i - \bm{r}_k \right|^2}\right), 
\label{eqfvv} \\
\bm{f}_{i}^{s} = f_{0} \left(\frac{1}{1 - r_i^2}- h\right)\bm{r}_i,
\label{eqfs}
\end{eqnarray}
where 
$h = (H_0 / 2H_{c2}) (R / \xi)^2$ 
is the dimensionless applied magnetic field, 
and $L$ is the number of vortices, or the vorticity. 
Our numerical approach is based on the Langevin dynamics algorithm, 
where the time integration of the equations of motion is performed 
in the presence of a random thermal force. 
The overdamped equation of motion (see, e.g., \cite{md01,md0157}) becomes: 
$
\eta \bm{v}_{i} \ = \ \bm{f}_{i} \ = \ \bm{f}_{i}^{vv} + \bm{f}_{i}^{vp} 
+ \bm{f}_{i}^{T} + \bm{f}_{i}^{d} + \bm{f}_{i}^{s}. 
$
Here
$\bm{f}_{i}^{d}$ 
is the driving force (Eq.~(\ref{eqfj})), 
$\bm{f}_{i}^{vp}$
is the force due to vortex-pin interaction, 
and
$\bm{f}_{i}^{T}$
is the thermal stochastic force,
obeying the condition: 
$
\langle \bm{f}_{\alpha, i}(t) \bm{f}_{\beta, j}(t') \rangle =
2\eta\,\delta_{\alpha\,\beta}\,\delta_{i\,j}\,\delta(t - t')k_BT,
$
where
Greek and italic indices refer to vector components and vortex labels,
$\eta$ is the viscosity which is set to unity. 
The ground state of the system is obtained by minimization of the London 
free energy, taking into account the vortex cores contribution \cite{CPB04}, 
with respect to: (i) different $L$, and (ii) different vortex configurations for 
the same $L$ (see Refs.~\cite{CBPB04,BCPB04}) by performing the 
stimulated annealing simulation (SAS) that simulates field-cooling experiments.

{\it Angular melting: Commensurability vs. the Lorentz force gradient.---} 
In a Corbino disk the gradient of the Lorentz force is maximum 
near the center, and thus 
one can expect that with increasing applied current the growing shear 
stress first breaks the vortex solid near the center and then, 
with further increasing the applied current, the melting process \cite{shear-melting}
propagates from the center to the peripheral regions.  
Indeed, this scenario of angular melting of the vortex solid 
in Corbino disks was observed in the experiment 
\cite{crabtree} 
and analyzed theoretically 
(e.g., \cite{paltiel,marchetti,miguel}). 
Similarly, in mesoscopic disks vortex shells first unlock 
near the center with increasing either applied current or 
temperature 
\cite{wecorbino}. 
Fig.~1 shows the angular velocity of different vortex shells 
$\omega_{i}$ 
for the configuration (1,5,12,18) as a function of the applied current
$I_{0}$. 
At low $I_{0}$, the whole vortex configuration rotates 
as a rigid body while at some critical value $I_{c12}$ the first  
inner shell splits off and starts to rotate separately with 
a {\it higher } angular velocity $\omega_{1}$. 
At a higher driving current $I_{c23}$, the Lorentz force gradient 
becomes sufficient for unlocking the second shell. 
For vortex configuration with a larger number of shells, the angular 
melting thus occurs gradually through the consequent unlocking of the 
shells from the center towards the periphery.

\begin{figure}[btp]
\begin{center}
\hspace*{-0.5cm}
\includegraphics*[width=7.5cm]{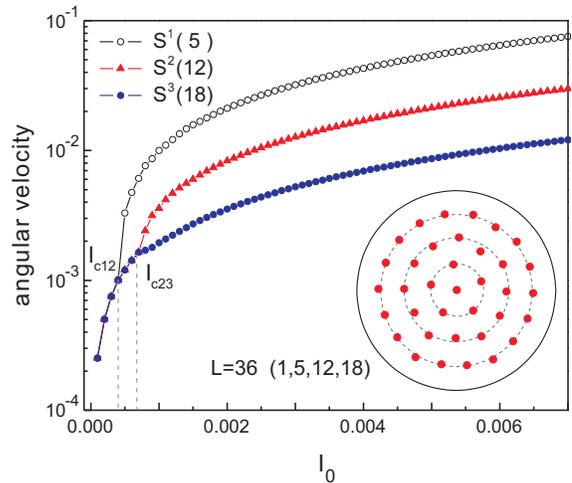}
\end{center}
\vspace{-0.5cm}
\caption{
(Color online)
The angular velocities $\omega_{i}$ of different vortex shells as a function 
of the external driving current $I_{0}$. 
The critical values $I_{c12}$ and $I_{c23}$ correspond to the unlock 
transition between the shells $S^{1}$-$S^{2}$ and $S^{2}$-$S^{3}$. 
The two-step angular melting starts at the center of the disk where the 
Lorentz force is maximum. 
The inset: The shell configuration (1,5,12,18) for the vorticity 
$L=36$ and $h=45$. 
The dashed grey circles showing different shells are guides for eye. 
}
\vspace{-0.5cm}
\end{figure}

It is easy to understand that the above ``classical'' behavior 
is the only possible scenario for angular melting of the vortex solid 
in large (macroscopic) Corbino disks. 
Indeed, in large disks (i.e., where the intervortex distance 
$a \ll R$) vortices form a uniform hexagonal Abrikosov lattice. 
It is clear that the shear modulus of such lattice is the same 
everywhere (except for the regions in the vicinity to the boundary). 
Thus, the strongest shear stress which occurs near the center of the 
disk breaks up the lattice first. 
However, in case of vortex shells in mesoscopic disks the situation 
turns out to be very different: the shear modulus becomes a function 
of the radius due to (in)commensurate number of vortices 
in different adjacent shells. 
(Moreover, the shear modulus can also depend on the angle if the 
circular symmetry of the vortex-shell configuration is broken by 
defects.) 
Thus we can expect that the angular melting in a system of vortex 
shells will happen in a different way if the shear modulus has a strong 
radial dependence, i.e., stronger than the difference in Lorentz force 
between adjacent shells. 

Because of the complexity of the system it is not possible to find 
the shear modulus analytically. 
However, we know that the shear modulus will be much 
smaller for incommensurate shells than for the commensurate case where 
``magic number configurations'' \cite{SPB95,BPB94} make the vortex 
system more rigid. 
If the inner shells are incommensurate, the Lorentz force gradient 
will facilitate the ``classical'' scenario of angular melting of 
the vortex solid and the unlocking of shells will occur at an even 
lower value of the driving current $I_{0}$. 
In contrast, if the outer shells are incommensurate, the friction 
between them can be so small that they {\it first} unlock, leading 
to unconventional angular melting, i.e., from the outer edge of 
the vortex solid inwards.

\begin{figure}[btp]
\begin{center}
\hspace*{-0.5cm}
\includegraphics*[width=8.0cm]{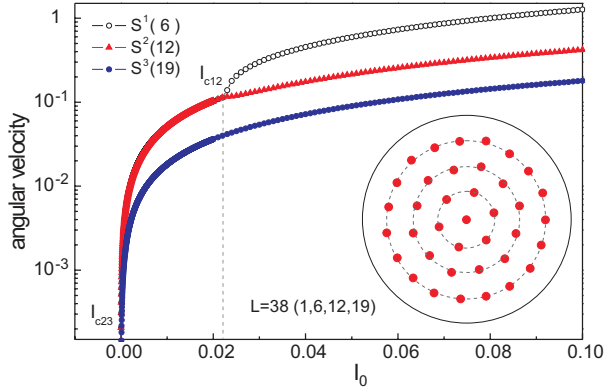}
\end{center}
\vspace{-0.5cm}
\caption{
(Color online)
Unconventional angular melting of vortex shells for the configuration (1,6,12,19): 
the two-step melting starts near the boundary where the Lorentz force is minimum. 
The inset shows the shell configuration (1,6,12,19) for the vorticity 
$L=38$ and $h=50$. 
}
\vspace{-0.5cm}
\end{figure}

We analyzed different ground-state and metastable vortex-shell 
configurations for different number of vortices $L$ and found 
different scenarios of angular melting in a mesoscopic Corbino disk. 
The most unusual way of melting is when it starts from the 
outermost shell and propagates {\it towards the center} of the disk, 
i.e., {\it opposite} to the ``classical'' way. 
An example of such unconventional melting is presented in Fig.~2. 
Note that it is realized for the ground-state energy 
configuration (1,6,12,19) for $L=38$. 
The circular-like incommensurate shell with 19 vortices easily 
slides with respect to the adjacent shell with 12 vortices 
even for very low driving currents $I>I_{c23}$ 
while the hard ``core'' (1,6,12) can sustain much stronger 
shear stress and the inner shell becomes unlocked at 
$I_{c12} \gg I_{c23}$ (Fig.~2).

\begin{figure}[btp]
\vspace*{-0.5cm}
\begin{center}
\hspace*{-0.5cm}
\includegraphics*[width=9.0cm]{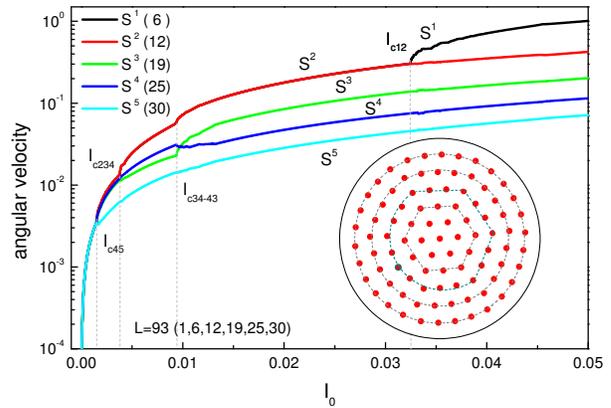}
\end{center}
\vspace{-0.5cm}
\vspace{-0.5cm}
\caption{
(Color online)
$\omega_{i}(I_{0})$ for the configuration (1,6,12,19,25,30). 
The system displays unconventional vortex dynamics in the region 
$I_{c234} \approx 0.004 \lesssim I_{0} \lesssim I_{c34-43} 
\approx 0.0095$ where the shell $S^{4}$ 
(with 25 vortices), 
that experiences a {\it weaker} Lorentz force, rotates {\it faster}  
than the 
adjacent shell $S^{3}$ (19 vortices) driven by a larger Lorentz force. 
The inset shows the shell configuration (1,6,12,19,25,30) for the 
vorticity $L=93$ and $h=109$. 
The $S^{3}$-shell consists of a hexagon-like part (CP) commensurate 
with the the hexagonal core (1,6,12) and a ``circular-like'' part (ICP) 
(with a defect) incommensurate with the core. 
}
\vspace{-0.5cm}
\end{figure}

{\it Unconventional dynamics of vortex shells.---} 
The interplay between (in)commensurability effects of vortex shells,  
on the one hand, and the inhomogeneous Lorentz force, on the other 
hand, can lead to even more striking and unexpected behavior. 
Thus, we demonstrate here that these effects can result in a very 
unusual (``unconventional'') dynamics of vortex shells in a mesoscopic 
Corbino disk. 
In particular, we found that in the region of the multistep transition 
from a ``rigid body'' rotation to individual rotation 
of vortex shells, an {\it inversion} of the angular velocities 
with respect to the gradient of the Lorentz force can occur. 
For unlocked shells we expect that, due to the $1/r$ dependence 
of the driving force, the angular velocity of the different shells 
is a {\it monotonically decreasing} function of the radius. 
This is clearly seen in Figs.~1 and 2. 
Very surprisingly, we found situations were this rule is broken 
as demonstrated in Fig.~3 for the $L=93$ configuration 
(1,6,12,19,25,30). 
The incommensurate 
shell $S^{4}$ with 25 vortices rotates {\it faster} than the adjacent 
shell $S^{3}$ (which is closer to the center than $S^{4}$) for 
$I_{c234} \approx 0.004 \lesssim I_{0} \lesssim I_{c34-43}  
\approx 0.0095$. 
This puzzling behavior is a consequence of the interplay 
between the Lorentz force, (in)commensurability effects and 
shell {\it defects}. 
For the above range of $I_{0}$, 
the vortex-shell configuration is only {\it partially} split. 
The strong commensurate inner-shell ``core'' rotates as a 
rigid body while incommensurate outer shells seem to rotate 
separately. 
However, their motion turns out to be {\it correlated} 
in a rather complex manner. 
Namely, the $S^{3}$-shell with 19 vortices has a ``defect'', 
i.e., one extra vortex 
(indicated by a blue circle in the inset of Fig.~3) 
which makes 
this shell incommensurate with the hexagon-like core (1,6,12). 
However, due to the inter-shell vortex-vortex interaction 
a part of this shell {\it squeezes} and adjusts itself to the 
core 
thus making the other part of the shell (i.e., incommensurate, 
containing the defect) {\it stretched}. 
The commensurate part (CP) of the $S^{3}$-shell 
(the hexagon-like part of shell $S^{3}$ -- see inset of Fig.~3) 
is locked to the core while the incommensurate part (ICP) 
(the ``circular'' part of shell $S^{3}$ -- see inset of Fig.~3)
is unlocked and slides with respect to the core. 
Thus the motion of the $S^{3}$-shell with respect to the core 
occurs in the form of a {\it compression-decompression wave} 
propagating in the direction {\it opposite} to the direction 
of the rotation. 
This backward wave motion reduces the average angular velocity 
of the $S^{3}$-shell by a factor of $18n/19$ where 
$n = 2\pi r_{S3}/\lambda_{CP-ICP}$ and $\lambda_{CP-ICP}$ is 
the wavelength of the CP-ICP compression-decompression wave 
\cite{cat}. 
At the same time, the incommensurate $S^{4}$-shell with 25 vortices, 
due to the friction with the CP of the $S^{3}$-shell, is 
{\it accelarated} by the faster rotation of the core. 
Note that this counter-intuitive behavior is only possible 
when a very delicate balance between the vortex-vortex 
interaction and the Lorentz force is established. 
When the latter becomes large enough (i.e., for $I_{0} \gtrsim 0.01$) 
the usual vortex-shell velocity distribution (i.e., monotonic 
decreasing) is recovered (see Fig.~3). 
Note that at $I_{0} \approx 0.01$ the angular velocity of the core 
also has a feature (kink) indicating the {\it complete} unlocking 
between the shells $S^{3}$, $S^{4}$ and the core (which itself breaks 
apart only for very large values of driving, 
$I_{0} > I_{c12} \approx 0.0325$).

{\it Conclusions.---} 
We found that the interplay between the (in)commensurability 
effects, the vortex-vortex interaction and the gradient Lorentz 
force in a mesoscopic Corbino disk can result in a very unusual 
and counter-intuitive dynamics of vortex shells. 
Thus we demonstrated: (i) unconventional angular melting 
when, with increasing driving current, the melting first occurs 
near the boundary where the shear stress is minimum 
and propagates towards the center, in an opposite way 
to the ``classical'' scenario of conventional angular melting 
in a Corbino disk; 
(ii) unconventional shell dynamics when a shell that experiences 
a weaker Lorentz force moves faster than the adjacent shell driven 
by a stronger Lorentz force. 
These examples of strongly non-linear dynamical behavior can be useful 
for understanding and prediction of dynamics of other complex interacting 
systems.

\smallskip

We thank Franco Nori and Fabio Marchesoni for useful discussions. 
This work was supported by the ``Odysseus'' program of the 
Flemish Government, FWO-Vl, and IAP. 
V.R.M. is funded by the EU Marie Curie project, 
Contract No. MIF1-CT-2006-040816.


\begin{references}
\bibitem{shear-melting} 
This ``angular melting'', or consequent unlocking of vortex shells, 
can be induced either by temperature or by shear (i.e., shear-induced 
plastic deformation). In the latter case, it is similar to the 
``shear-induced melting transition'' in colloidal crystals.
\bibitem{cat} 
Note that this kind of motion resembles the motion of a caterpillar 
whose part of the body ``locks'' to the surface while other part 
moves forward (decompression), and then it ``locks'' and pulls 
the other part (compression).
\end{references}
\end{document}